\documentclass[aps,prb,showpacs,twocolumn]{revtex4}
\usepackage{epsfig}
\usepackage{amsmath}

\begin{document}

\title{On-Site Interaction Effects on Localization : Dominance of Non-Universal
Contributions}
\author{Moshe Goldstein}
\author{Richard Berkovits}
\affiliation{The Minerva Center, Department of Physics, Bar-Ilan University,
  Ramat-Gan 52900, Israel}

\begin{abstract}
The influence of on-site (Hubbard) electron-electron interaction
on disorder-induced localization is studied in order to clarify
the role of electronic spin. The motivation is based on the recent
experimental indications of a ``metal-insulator'' transition in
two dimensional systems. We use both analytical and numerical
techniques, addressing the limit of weak short-range interaction.
The analytical calculation is based on Random Matrix Theory (RMT).
It is found that although RMT gives a qualitative explanation of
the numerical results, it is quantitatively incorrect. This is  due to
an exact cancellation of short range and long range correlations in RMT,
which does not occur in the non-universal corrections to RMT. An
estimate for these contributions is given.
\end{abstract}


\maketitle

\section{Introduction}
The question considered in this paper is whether electron-electron interaction
can reduce disorder-induced localization, thus enabling metallic behavior in
two dimensional disordered systems.

The common view about the subject in the last 20 years has been based
on the well known scaling theory of localization \cite{gang79},
according to which two dimensional systems will always be
localized (i.e., insulating), no matter how weak the disorder is.
Although the original scaling theory did not take interactions
into account, it was shown that weak interaction (i.e., high
electron-density) does not affect its results \cite{altshuler80}. On
the other hand, in the limit of very strong interaction (i.e.,
very dilute systems) it is known that the electron liquid freezes
into a Wigner lattice, which is pinned by disorder and therefore
insulating \cite{tanatar89}. All these results have lead to the
opinion that the repulsion between electrons can only further
decrease the conductance, so that all two dimensional systems will
show insulating behavior, regardless of the strength of
interaction between the electrons.

A series of experiments performed in the last few years showed
that even though in the limits of both very dense and very dilute
systems we get the expected insulating behavior, for intermediate
values of density (corresponding to $r_s$ between 4 and 40, where
$r_s$ is the average inter-electron distance measured in the units
of the Bohr radius) metallic-like temperature dependence is
found\cite{review01}. The transition from an insulating behavior
to a metallic one as the density decreases was entitled ``Two
Dimensional Metal-Insulator Transition'' (2DMIT). An important
feature of these systems is that an application of an in-plane
magnetic field, (which cannot affect the electrons' orbital motion
but can direct their spins) reduces the conductance in  the
metallic regime, until for high enough magnetic fields the
conductance saturates as a function of the field, and the systems
show the expected insulating behavior. This saturation field was
estimated to be the field of full alignment of all the spins.

These results arouse much interest and many ideas where suggested
for their explanation. A debate started in the question of whether
there is really a metallic behavior and a phase transition,
probably caused by electron-electron interaction \cite{finkelstein84}; or the
system is really insulating, but the experimentally accessible temperatures
are high enough to exhibit temperature dependent scattering, thus
causing the apparent metallic behavior \cite{aleiner01}.

Analytical \cite{finkelstein84} and numerical \cite{berkovits01}
calculations have shown that, as expected, for spinless electrons
repulsion can only further localize the electrons, and does not
lead to a metal-insulator transition. However, when taking spin
into account, the situation is still unclear \cite{finkelstein84}.
In a recent numerical exact-diagonalization study
\cite{berkovits02}, an Anderson model with both long range Coulomb
interaction and short range Hubbard interaction was considered. It
was shown that the Coulomb interaction, existing between any two
electrons regardless of their spin, can only increase
localization. On the other hand, not-too-strong Hubbard
interaction were seen to cause delocalization (Strong Hubbard
interaction will lead to a Mott-Hubbard insulator). Since this
interaction exists only between electrons with opposite spins, its
effect is decreased by an in-plane magnetic field, and disappears
when all the spins are aligned. This dependence of localization on
interaction-strength and in-plane magnetic field thus mimics, at
least qualitatively, the experimentally observed phenomena.
Similar results were obtained recently using Quantum Monte-Carlo
methods \cite{dutch03}.

In this paper we wish to study further the
weak short-range interaction regime, in which interaction-induced
delocalization was observed. We will first address the problem
analytically, using a Random Matrix Theory (RMT) approach
\cite{mehta91}, and then compare it to numerical simulations on an
Anderson model. It will be shown that RMT can give only a
qualitative but not a quantitative explanation for the numerical
results, since RMT does not take into account non-universal
correlations existing between wave functions in the diffusive
regime. An estimate for the effect's order of magnitude and its
dependence on the parameters of the system in the diffusive regime
will be given.

\section{Analytical Results - Random Matrix Theory}

We will consider an Anderson Hamiltonian with on-site Hubbard
interaction :
\begin{eqnarray} \label{eqn:hamiltonian}
{\hat H} =
\displaystyle \sum_{s;\sigma}\epsilon_{s} {\hat n}_{s;\sigma}
- t \negthickspace \displaystyle \sum_{<s,s\prime>;\sigma}
{\hat a}^{\dagger}_{s;\sigma}{\hat a}_{s\prime;\sigma}
+ U_H  \negthickspace \displaystyle \sum_{s} {\hat n}_{s;\uparrow}
{\hat n}_{s;\downarrow} ,
\end{eqnarray}
where ${\hat a}^{\dagger}_{s;\sigma}$, ${\hat a}_{s\prime;\sigma}$
and ${\hat n}_{s;\sigma}$ denote electron creation, annihilation
and number operators, respectively, for a state on site $s$ with
spin projection $\sigma$ on some axis. The first term is a random
on-site potential, where $\epsilon_s$ is chosen randomly from the
range [-W/2,W/2]; the second is the hopping or kinetic term, where
the sum is over nearest-neighbor sites $s$, $s\prime$ and $t$ is
an overlap integral; the third is the Hubbard term, the
electrostatic interaction between two electrons in the same site
(which must have opposite spins), whose strength is determined by
the parameter $U_H$.

To quantify localization, we will calculate the Inverse
Participation Ratio (IPR), defined by $P^{-1}=\sum_{s}
{\left|\psi(s)\right|}^4$. This quantity is of order 1 for
localized states, and of order $N^{-1}$ for delocalized states,
where $N$ is the  number of lattice sites. The IPR thus decreases
when the single-particle wave function $\psi$ becomes less
localized, and gives us an estimation for the changes in the
conductance of the system.

We assume here that without interaction the single electron energies and
eigenvectors distributions for the ensemble of Anderson Hamiltonians are
described by the corresponding distributions for an ensemble of Gaussian real
symmetric matrices, i.e., the Gaussian Orthogonal Ensemble (GOE). This ensemble
is defined by the well known distribution \cite{mehta91}:
\begin{equation} \label{eqn:rmt}
P(H)\mu(H)=\exp(-\frac{\beta}{4{\lambda}^2}Tr(H^2))\mu(H) ,
\end{equation}
where $\beta=1$, $\lambda$ is a constant energy parameter, and $\mu(H)$ is a
suitable measure. The eigenvectors are then a set of random orthogonal real
normalized vectors. The average IPR without interaction for an electron
in the n-th level with spin $\sigma$ is thus \cite{ullah63}
\begin{equation} \label{eqn:rmtipr}
P_n^{-1}= \displaystyle \sum_{s} \langle\langle
{\left|\psi^{(0)}_{n;\sigma}(s)\right|}^4 \rangle\rangle=\frac{3}{N+2} ,
\end{equation}
where the superscript (0) denotes the state without interaction, and double
angular brackets denote ensemble average.

Now we add a weak Hubbard interaction, treating it in a self consistent way to
first order in perturbation theory. Thus, the effect of spin-down electrons
on the electrons with spin up will be the following effective potential (since
only electrons with different spins interact, we have no exchange term):
\begin{eqnarray}\label{eqn:potential}
{\hat V} = U_H \displaystyle \sum_{s}
{\left|\psi^{(0)}_{m;\downarrow}(s)\right|}^2 {\hat n}_{s;\uparrow} .
\end{eqnarray}

According to the familiar first order perturbation theory, the first order
change in the IPR of a spin-up electron in the n-th state due to its
interaction with a spin-down electron in the m-th state is :
\begin{eqnarray}\label{eqn:dipr}
& \Delta_m P_n^{-1} \sim \nonumber \\
& 4 U_H \negthickspace \displaystyle \sum_{\substack{l \neq n \\ s,s\prime}}
\langle\langle \frac{ {\left(\psi^{(0)}_{m;\downarrow}(s\prime)\right)}^2
\psi^{(0)}_{n;\uparrow}(s\prime) \psi^{(0)}_{l;\uparrow}(s\prime)
{\left(\psi^{(0)}_{n;\uparrow}(s)\right)}^3 \psi^{(0)}_{l;\uparrow}(s) }
{E^{(0)}_n-E^{(0)}_l} \rangle\rangle . \nonumber \\
\end{eqnarray}
Since the wave functions can be chosen to be real due to time
reversal symmetry, we omitted absolute value and complex conjugate
notations in this and the following expressions.

According to RMT, the eigenvectors distribution is independent
of the eigenvalues distribution, so we can separate the averages of the
numerator and denominator in the above expression.

\begin{table*}

\begin{tabular}{|c|c|c|} \cline{2-3}
\multicolumn{1}{c|}{} &
{\large $s=s\prime$ } & {\large $s \neq s\prime$ } \\ \hline
{\large $l\neq m\neq n$ } & {\Large $\frac{3}{N(N+2)(N+4)(N+6)}$ }  &
{\Large $-\frac{3(N+3)}{(N-1)N(N+1)(N+2)(N+4)(N+6)}$ } \\ \hline
{\large $l=m \neq n$ } & {\Large $\frac{9}{N(N+2)(N+4)(N+6)}$ } &
{\Large $-\frac{9(N+3)}{(N-1)N(N+1)(N+2)(N+4)(N+6)}$ } \\ \hline
{\large $m=n \neq l$ } & {\Large $\frac{15}{N(N+2)(N+4)(N+6)}$ } &
{\Large $-\frac{9}{(N-1)N(N+2)(N+4)(N+6)}$ } \\ \hline
\end{tabular}

\caption{\label{tbl:avr}Values of the average of the numerator in
Eq.~(\ref{eqn:dipr}) for all the possible combinations of level numbers
l, m, n and sites s, s'.}
\end{table*}

As for the average of the numerator, its value can be found in the
literature \cite{ullah63, gorin02}, and the results are summarized
in Table~\ref{tbl:avr}. We note that when $s=s\prime$ we have an
average of even powers of wave functions at different sites, which
is expected to be positive and vary as $N^{-4}$, since we have
eight wave function values in the expression, each of which goes
as $N^{-1/2}$. On the other hand, when $s \neq s\prime$, it may
appear at first glance that since we have an average of odd powers
of values of wave functions at different sites, which are
uncorrelated, we should get zero. However, we get in this case a
nonzero negative value, going as $N^{-5}$. This result is due to
correlations resulting from the orthogonality requirement on the
eigenvectors.

To understand this, we may note that squaring the orthogonality relation
$\sum_{s}\psi_{j}(s) \psi_{k}(s)=0 $ for $j \neq k$ and averaging, using the
known result \cite{ullah63}
\begin{eqnarray} \label{eqn:example1}
\langle\langle {\left( \psi_{j}(s) \right)}^2 {\left( \psi_{k}(s) \right)}^2
\rangle\rangle = \frac{1}{N(N+2)} ,
\end{eqnarray}
we find that
\begin{eqnarray} \label{eqn:example2}
\langle\langle \psi
_{j}(s) \psi_{j}(s\prime) \psi_{k}(s) \psi_{k}(s\prime)
\rangle\rangle = -\frac{1}{(N-1)N(N+2)},
\end{eqnarray}
for $s \neq s\prime$, i.e., if two different wave functions have the same sign
on one site, from orthogonality they will tend to have opposite signs on
another site and vice versa, hence the above nonzero negative average.

As for the average value of the energy denominator in Eq.~(\ref{eqn:dipr}),
in principle it might be possible to calculate its value using RMT. However,
to estimate the leading order we will assume the spectrum is composed of
equidistant levels, with mean level spacing $\Delta$.

Combining all those results together, we get, to the leading order in $N$,
the following result for the change in the IPR of a spin-up electron in the
n-th level due to its interaction with a spin-down electron in the m-th level:
\begin{eqnarray*}
\Delta_m P_n^{-1} =&&
\end{eqnarray*}
\begin{eqnarray}
& \mbox{\Large $-\frac{24}{N^3} \frac{U_H}{\Delta}$}
\left( \Phi(N-n)-\Phi(n-1) \right), & m=n; \nonumber \\
& \mbox{\Large $\frac{24}{N^4} \frac{U_H}{\Delta}$}
\left( \Phi(N-n)-\Phi(n-1) + \mbox{\Large $\frac{2}{m-n}$ } \right),
& m\neq n; \nonumber \\
\end{eqnarray}
where $\Phi(k)$ is defined by:
\begin{eqnarray*}
\Phi(n) = \displaystyle \sum_{k=1}^{n} \frac{1}{k} .
\end{eqnarray*}
We observe that for $m=n$ the correction is always negative (for n in the
lower half of the band), i.e., the interaction between two electrons in the
same state tends to delocalize them, which is the only way to reduce their
mutual interaction energy. For $m\neq n$ the correction will usually be
positive, i.e., electrons in different levels repulse each other, resulting
in further localization. As can be
expected, the former effect is larger than the latter, due to the identity
of the two interacting electrons' wave functions in the former case. However,
the order $N$ difference between the case $m=n$ and the case $m \neq n$ is
caused by an excat cancellation of the leading order dependence on $N$ between
the single short range ($s=s\prime$) term and all the $N-1$ long range
($s \neq s\prime$) terms in the latter case, which doesn't occur in the
former. We will see below that this cancellation, together with the negative
sign of the result for $m \neq n$, is correct only in RMT.

Thus, if the lowest $n_{\downarrow}$ levels are occupied by spin-down
electrons, the total change in the IPR of a spin-up electron
in the n-th level is:
\begin{eqnarray*}
\Delta P_n^{-1} =&&
\end{eqnarray*}
\begin{eqnarray} \label{eqn:final}
& \mbox{\Large $-\frac{24}{N^3} \frac{U_H}{\Delta}$}
\left( 1- \mbox {\Large $\frac{n_{\downarrow}-1}{N}$} \right)
\left( \Phi(N-n)-\Phi(n-1) \right) & \nonumber\\
& + \mbox{\Large $\frac{48}{N^4} \frac{U_H}{\Delta}$}
\left( \Phi(n_{\downarrow}-n)-\Phi(n-1)\right) , &
n\le n_{\downarrow} ; \nonumber \\ \nonumber \\ \nonumber \\
& \mbox{\Large $\frac{24n_{\downarrow}}{N^4} \frac{U_H}{\Delta}$}
\left( \Phi(N-n)-\Phi(n-1) \right) & \nonumber\\
& - \mbox{\Large $\frac{48}{N^4} \frac{U_H}{\Delta}$}
\left( \Phi(n-1)-\Phi(n-n_{\downarrow}-1)\right) , &
n > n_{\downarrow} . \nonumber \\
\end{eqnarray}
The main features in the behavior of $\Delta P_n^{-1}$ are as follows~:
For $n\le n_\downarrow$ the negative contribution of the spin-down
electron at the same level n as the affected spin-up electron dominates the
usually positive contribution of the other spin-down electrons.
Therefore, $\Delta P_n^{-1}$ is negative, but decreases in absolute value when
$n_\downarrow$ increases. For $n>n_\downarrow$, there are spin-down
electrons only in levels different from n, thus $\Delta P_n^{-1}$ is positive
and increases when $n_\downarrow$ increases.
At $n=n_\downarrow$ there is a discontinuous jump of $\Delta P_n^{-1}$. In both
cases, since $\Delta \sim N^{-1}$ in real systems (although not in RMT),
the effect is of order $N^{-2}$, if we keep
the concentration of spin-down electrons constant. (We neglect here the
logarithmic factor coming from the function $\Phi(n)$). A plot of these
formulas will be shown in the next section, where these expressions will be
compared to numerical results.

\begin{figure*} \centering
\epsfxsize15cm\epsfbox{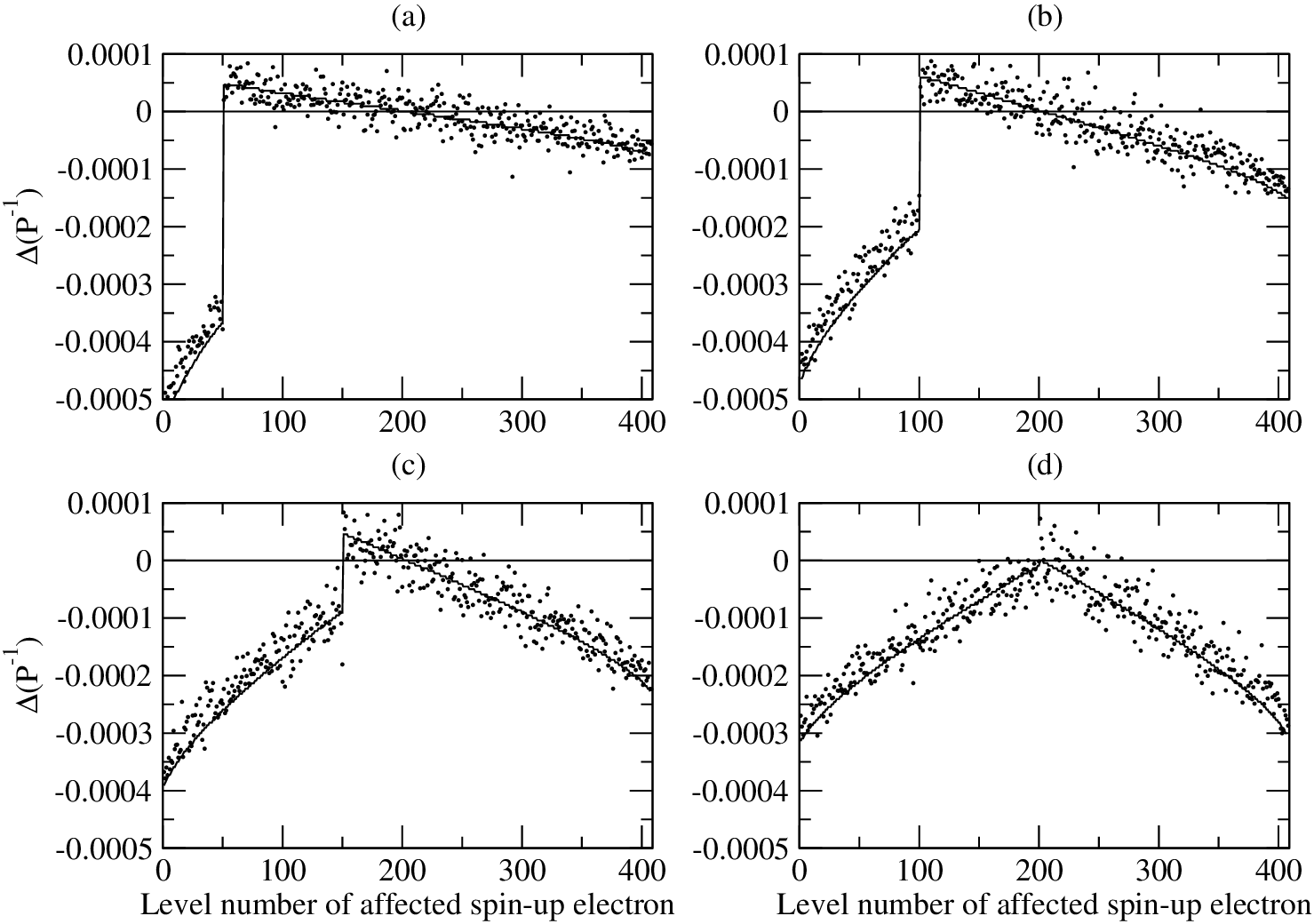}\caption{\label{fig:rmt_dipr}
Change in the IPR of a spin-up electron due to its interaction with spin-down
electrons, according to RMT. The change is plotted as a
function of the level number of the affected spin-up electron for different
numbers of spin-down electrons: (a) $n_\downarrow=50$; (b) $n_\downarrow=100$;
(c) $n_\downarrow=150$; (d) $n_\downarrow=200$.
In all the graphs the line indicates the theoretical formula, while the dots
indicate the numerical results.  The numerical results are averages over an
ensemble of $5\times 10^4$ realizations of $408\times 408$ RMT Hamiltonians.
The estimated error approximately equals the
width of the numerical results. Further parameters are given in the text.}
\end{figure*}

\section{Numerical Results}
In this section we will examine results of numerical calculations
and compare them to the analytical results discussed above. Two
model Hamiltonians will be considered : an RMT Hamiltonian and an
Anderson Hamiltonian. It will be shown that their results
differ by an order of magnitude as well as in other
characteristics. The theoretical predictions will be shown to
agree with the former but not with the latter, and reasons for the
discrepancy will be given.

\subsection{Random Matrix Hamiltonian}
We will first consider the change in the IPR for a true RMT Hamiltonian.
Since we consider here only the  weak interaction regime, instead of solving
the exact many-body problem we simply diagonalize first the Hamiltonian
without interaction, and then use the wave functions to construct the
effective potential, given in Eq.~(\ref{eqn:potential}). This potential is
then used to calculate the wave functions and the IPR with
interaction. The applicability of this one loop Hartree-Fock approximation
is justified by the fact that the change in $P_n^{-1}$ was found to be linear
in $U_H$, as required.

The matrix size chosen was $408\times 408$, and the elements were chosen
according to the distribution law in Eq.~(\ref{eqn:rmt}). We have chosen
$\lambda=0.1t$, so that the mean level spacing is $\Delta=0.0196t$,
approximately equal to the spacing in the Anderson Hamiltonian,
Eq.~(\ref{eqn:hamiltonian}), used in the next section (0.022t to 0.025t for W
between 2.0t and 4.0t). The interaction strength $U_H$ was taken as 1.0t. The
calculated quantities were averaged over an ensemble of $5\times 10^4$
different realizations.

The numerical results for the change in the IPR vs. the level number of the
affected spin-up electron due to its interaction with different numbers of
spin-down electrons, are shown in Fig.~\ref{fig:rmt_dipr}, together with the
theoretical formula, Eq.~(\ref{eqn:final}). The theoretical formula was
corrected, taking into account that the mean level spacing is not constant
across the spectrum, but varies according to the semicircle law \cite{mehta91},
\begin{equation}
\frac{1}{\Delta(E)}=\rho(E)=
\frac{1}{2\pi{\lambda}^2\beta} \sqrt{4{\lambda}^2\beta N-E^2},
\end{equation}
where $\rho(E)$ is the density of states.

As can be seen, there is a good agreement between the numerical
and the theoretical results. All the main features discussed at the end of
the previous section can be clearly seen in the numerical data.

\subsection{Anderson Hamiltonian}
Now we will discuss the changes in the IPR for the Anderson
Hamiltonian given in Eq.~(\ref{eqn:hamiltonian}). The calculation
was performed in the same method as was used for the random matrix
Hamiltonian (i.e., one-loop Hartree-Fock approximation).

\begin{figure} \centering
\epsfxsize8cm \epsfbox{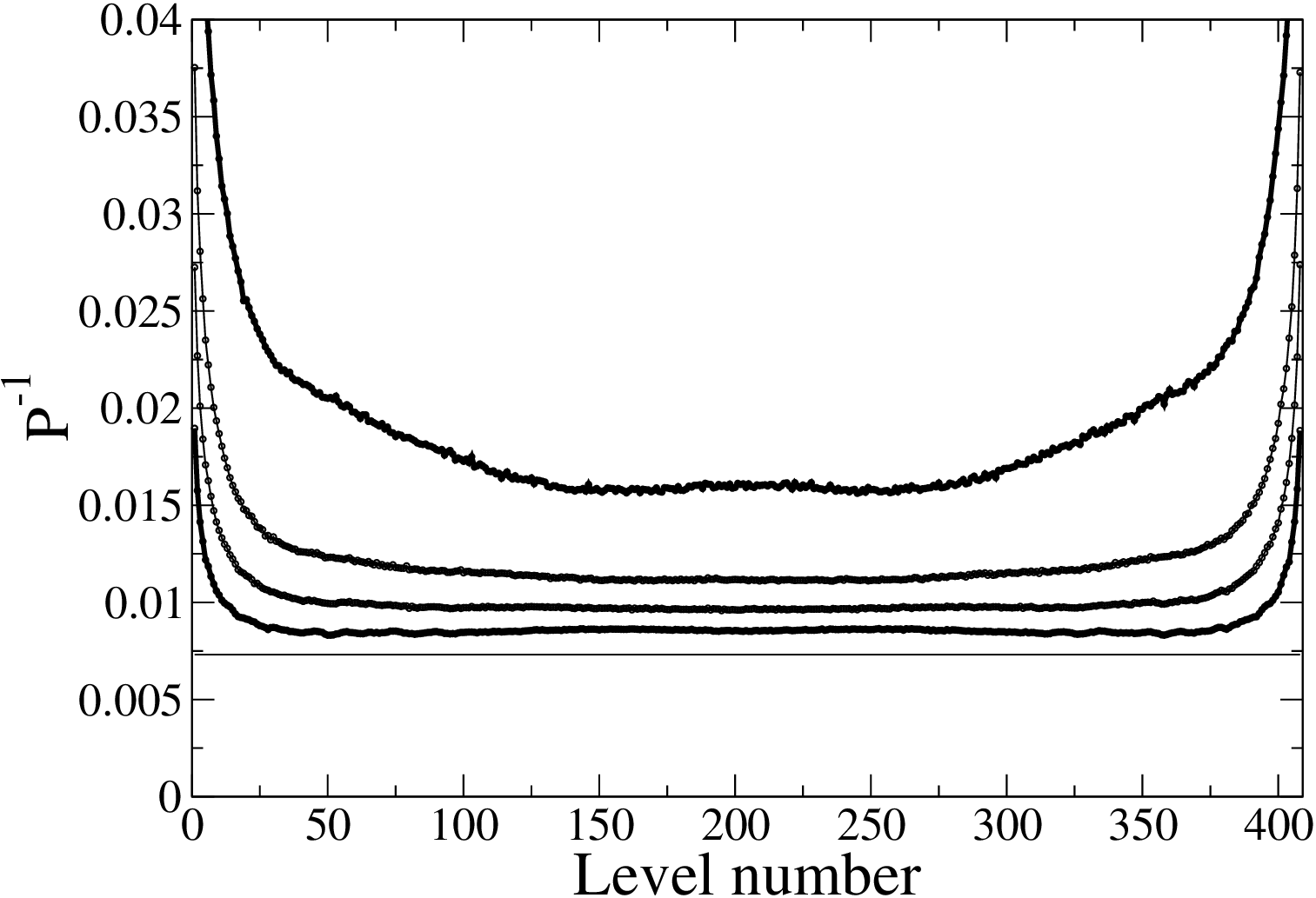}\caption{\label{fig:an_ipr}
The IPR for non-interacting electrons in the Anderson model. The IPR is
plotted as a function of the level number. The lowest curve shows the RMT
value, while the other ones are the Anderson model results for W=2.0t, W=2.5t,
W=3.0t and W=4.0t, from lower to upper, respectively. The results are averages
over an ensemble of $10^4$ realizations of systems on a $17 \times 24$ sites
lattice. The estimated error approximately equals the width of the
numerical results. Further parameters are given in the text.}
\end{figure}

\begin{figure*} \centering
\epsfxsize15cm\epsfbox{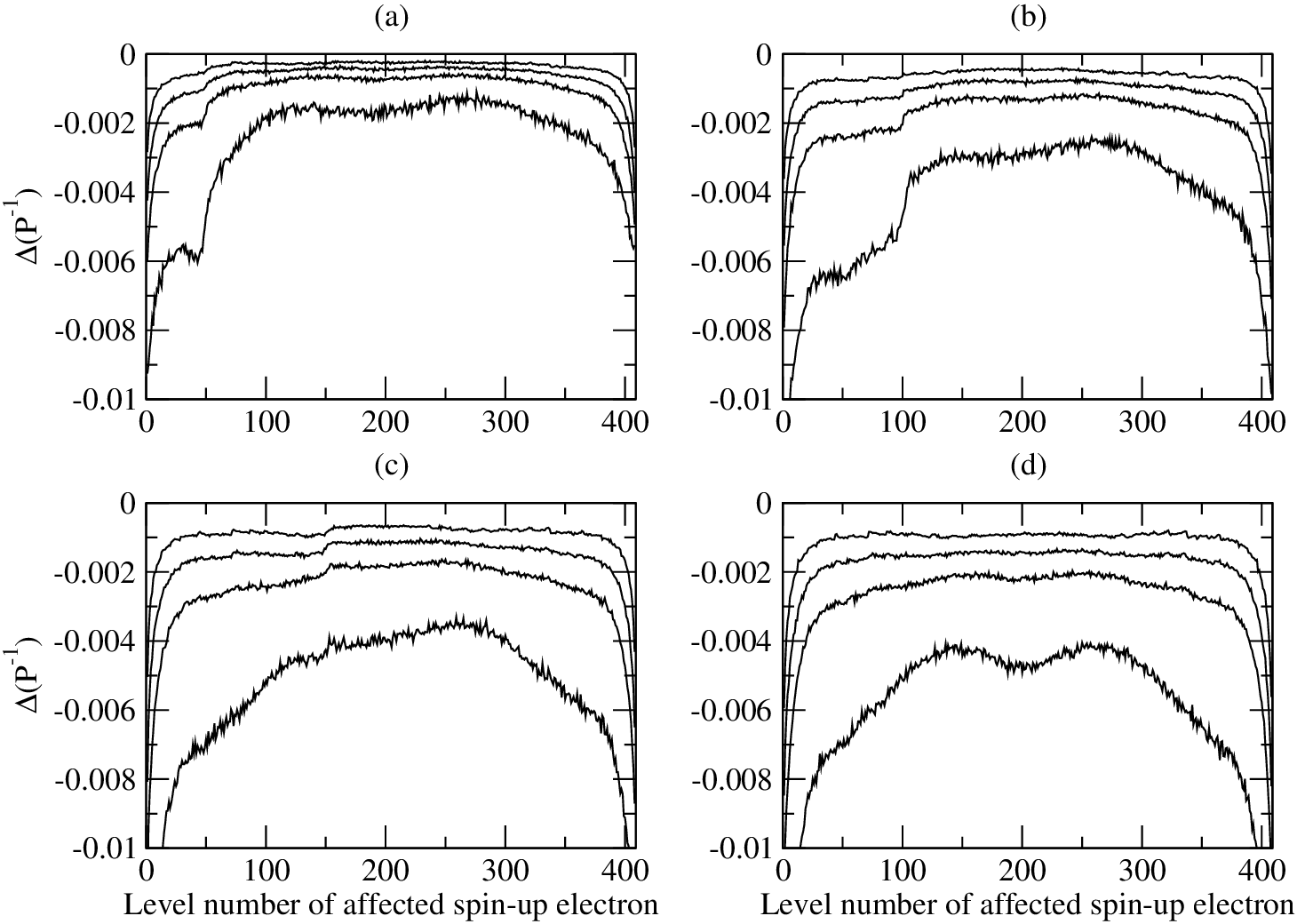}\caption{\label{fig:an_dipr}
Change in the IPR of a spin-up electron due to its interaction with spin-down
electrons in the Anderson model. The
change is plotted as a function of the level number of the affected spin-up
electron for different numbers of spin-down electrons: (a) $n_\downarrow=50$;
(b) $n_\downarrow=100$; (c) $n_\downarrow=150$; (d) $n_\downarrow=200$.
In all the graphs the curves correspond to W=4.0t, W=3.0t, W=2.5t and W=2.0t,
from lower to upper, respectively. The results are averages
over an ensemble of $10^4$ realizations of systems on a $17 \times 24$ sites
lattice. The estimated error approximately equals the width of the
numerical results. Further parameters are given in the text.}
\end{figure*}

\begin{figure*} \centering
\epsfxsize15cm\epsfbox{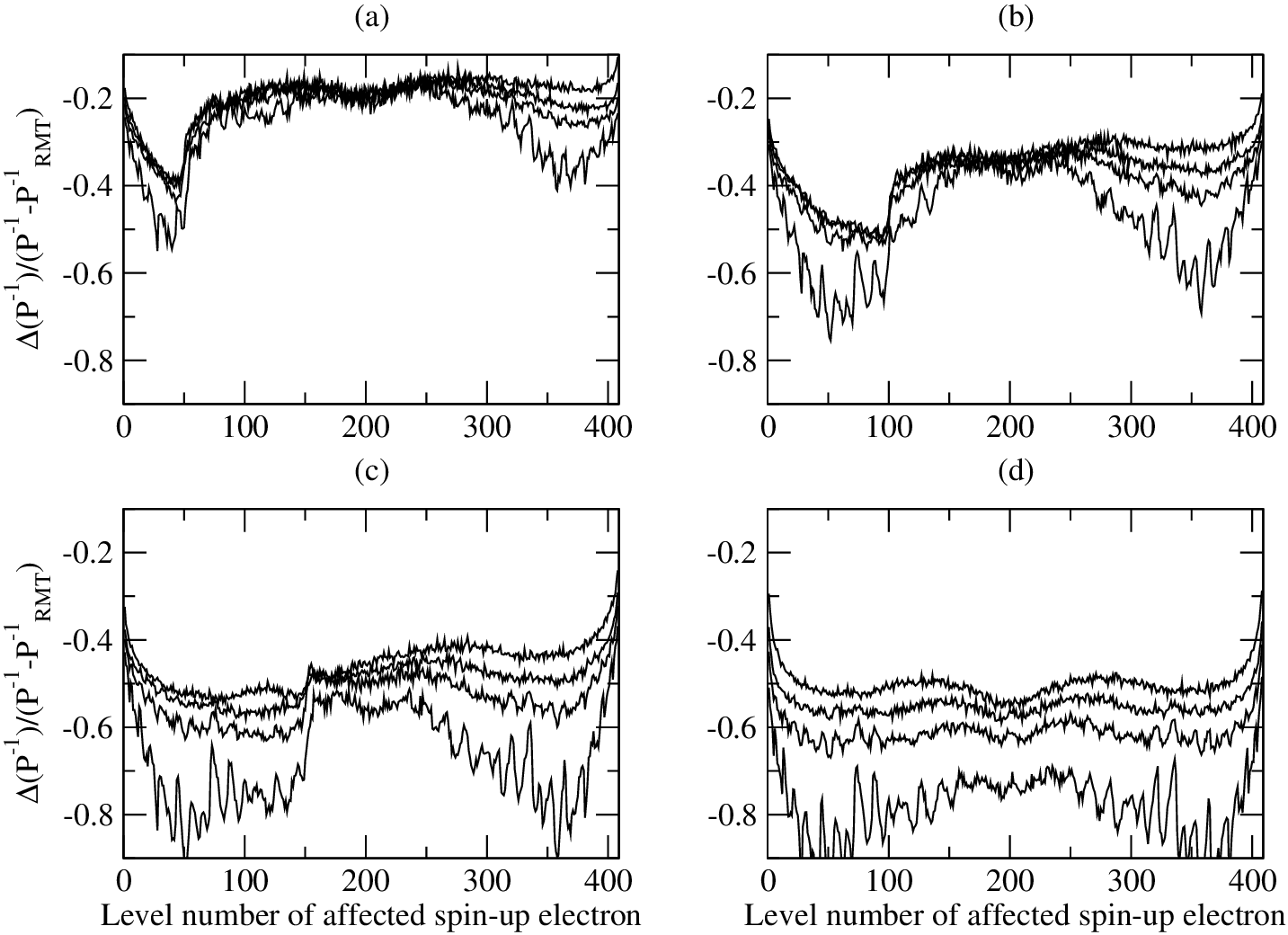}\caption{\label{fig:an_diprd}
Ratio between the change in the IPR of a spin-up electron due to its
interaction with spin-down electrons in the Anderson model and the
non-universal part of the IPR without interaction.
The ratio is plotted as a function of the level number of the affected spin-up
electron for different numbers of spin-down electrons: (a) $n_\downarrow=50$;
(b) $n_\downarrow=100$; (c) $n_\downarrow=150$; (d) $n_\downarrow=200$.
In all the graphs the curves correspond to W=2.0t, W=2.5t, W=3.0t and W=4.0t,
from lower to upper, respectively. The results are averages
over an ensemble of $10^4$ realizations of systems on a $17 \times 24$ sites
lattice. The estimated error approximately equals the width of the
numerical results. Further parameters are given in the text.}
\end{figure*}

\begin{figure*} \centering
\epsfxsize15cm\epsfbox{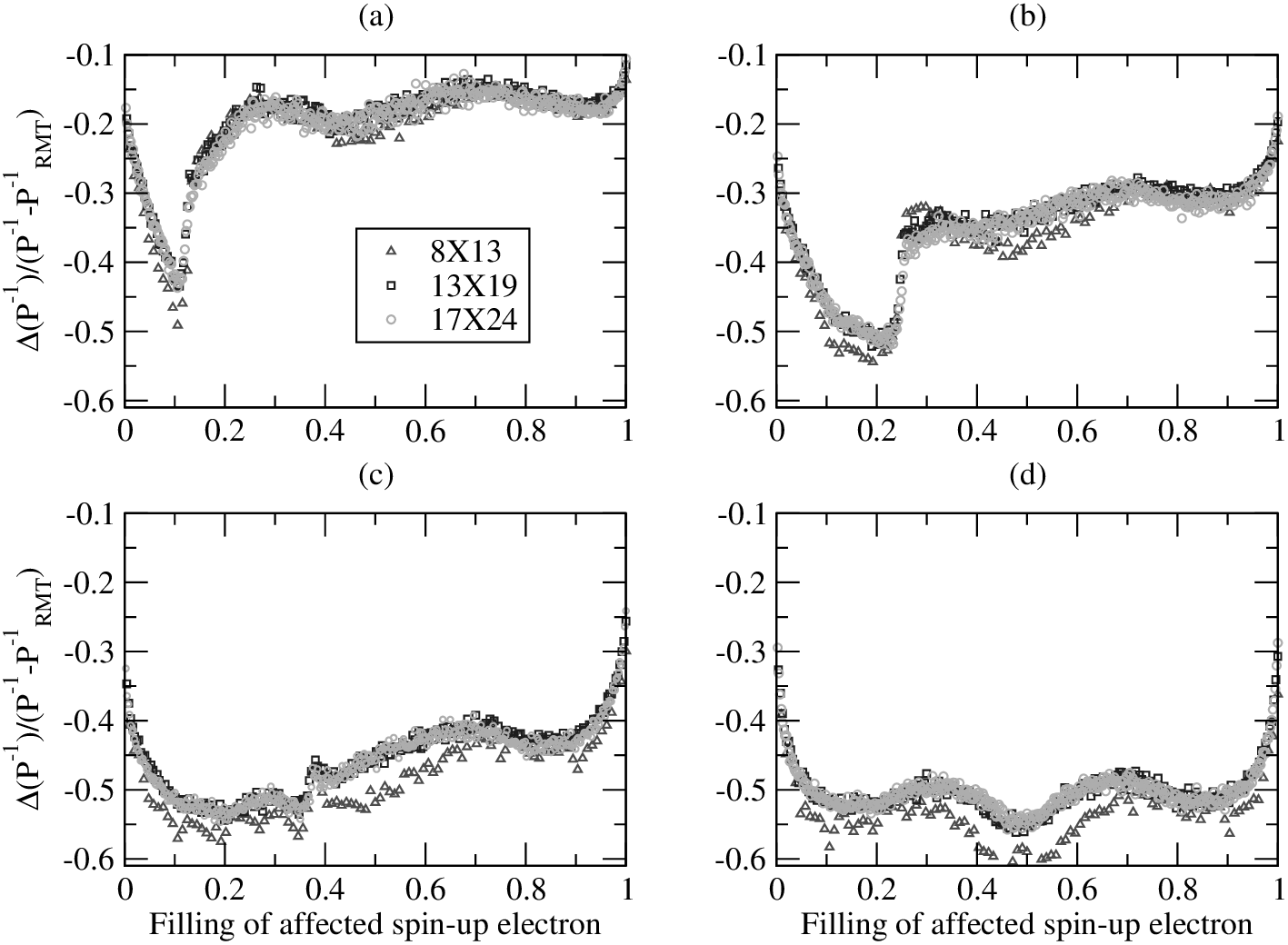}\caption{\label{fig:an_diprn}
Ratio between the change in the IPR of a spin-up electron due to its
interaction with spin-down electrons in the Anderson model and the
non-universal part of the IPR without interaction. The ratio is plotted
as a function of the filling of the affected spin-up electron (i.e., the
ratio of the number of spin-up electrons and the number of lattice
sites) for different fillings of spin-down electrons:
(a) $\nu_\downarrow \approx 1/8$; (b) $\nu_\downarrow \approx 1/4$;
(c) $\nu_\downarrow \approx 3/8$; (d) $\nu_\downarrow \approx 1/2$.
In each graph we use three different lattice sizes -- $8\times 13$,
$13\times19$, $17\times 24$, but a constant value of disorder, W=4.0.
The results are averages over an ensemble of $10^4$ realizations.
The estimated error approximately equals the width of the
numerical results. Further parameters are given in the text.}
\end{figure*}

We have chosen a $17\times 24$ lattice, corresponding to a $408\times 408$
matrix. As for the RMT calculations, we took $U_H=1.0t$, while four values of
disorder were used -- W=2.0t, W=2.5t, W=3.0t and W=4.0t. The results were
averaged over $10^4$ realizations of disorder.

First, in Fig.~\ref{fig:an_ipr}, the value of the IPR without
interaction is shown for the four values of disorder, as well as
the RMT value, Eq.~(\ref{eqn:rmtipr}). We can see a difference
here, as the Anderson model gives higher values (more localized)
of the IPR than RMT. The effect is caused by non-universal (i.e.,
beyond RMT) corrections to the IPR and is more pronounced for
higher disorder. The corrections for the IPR were calculated using
supersymmetry techniques \cite{prigodin98}, resulting in
$P^{-1}-P_{RMT}^{-1} \sim g^{-1} N^{-1}$(where $g$ is the
dimensionless conductance). We can also see, as expected, that the
levels near the band edge have higher IPR, and are thus more
localized, than levels near the center of the band.

Now we move to interaction effects in the Anderson model. The
results are shown in Fig.~\ref{fig:an_dipr}, with the same
occupation numbers as those chosen in the previous RMT
calculations, for the four values of the disorder. As in RMT, the
change in the IPR is negative for $n\le n_\downarrow$ and changes
sharply (though not discontinuously) at $n=n_\downarrow$.
Nevertheless, it doesn't change its sign there. Moreover, the
change in the IPR is larger by about an order of magnitude than
the one found from RMT. Also, even in the range $n \le
n_\downarrow$, it increases in absolute value, rather than
decreases, when $n_\downarrow$ increases. All this is in contrast
with Eq.~(\ref{eqn:final}) and the discussion following it.

Another point is that the effect increases with disorder. This is
seen by comparing $\Delta P_n^{-1}$ for the same level n but
different values of W; or by observing that, for the same value of
W, levels near the band edge, which are more localized, show
larger $\Delta P_n^{-1}$.

The reason for these differences is the above mentioned cancellation
between long range and short range wave-function correlations in RMT. As
has been seen in our RMT calculations (Table~\ref{tbl:avr}), the average
of wave functions product appearing in the numerator of Eq.~(\ref{eqn:dipr}),
is of order $N^{-4}$ and positive when the two sites considered
coincide, but are only of order $N^{-5}$ and negative when the
sites are different. Since there are $N-1$ terms of the latter type
for each term of the former type, their total contributions are
of the same order but their signs are opposite. Due to the equallity of the
numerical coefficients of the two types of terms when the interacting
electrons are in different levels, they cancel out exactly to the leading
order in $N$, leaving behind a small negative term, of order $N^{-5}$.
Therefore, in RMT interaction between electrons in different levels
increases their localization, opposite to the situation for
electrons in the same level. From this followed the decrease in
the absolute value of $\Delta P_n^{-1}$ as $n_\downarrow$
increases in the range $n \le n_\downarrow$, its positive value
for $n>n_\downarrow$, and the overall $N^{-2}$ dependence of the
effect for constant density of spin-up electrons.

All this is correct when $g$ is infinite. For finite $g$
there exist non-universal corrections to the wave-function averages.
Those corrections were not calculated before for the averages required
here, but their behavior can be conjectured from known corrections for
simpler averages (like those in Eqs.~(\ref{eqn:example1},
\ref{eqn:example2}) \cite{mirlin01}). We may expect them to
have the same $N$ dependence and sign as the RMT value, but to be smaller by a
factor of $g$. The corrections for the short range ($s=s\prime$) terms and
long range ($s \neq s\prime$) terms will not, in general, have equal numerical
coefficients, even when the interacting electrons are in different levels.
Hence, after summation over $s\prime$ we are left with
an order $g^{-1}N^{-4}$ contribution instead of the order $N^{-5}$
contribution in RMT. For this reason, although the
non-universal corrections are of order $g^{-1}$, for most of the
averaged terms they are about $N$ times larger, so they will
determine both the magnitude and the sign of the
interaction-induced change in the IPR. Since the corrections
for $s=s\prime$ will, in general, have a long range part, persisting
for $s \neq s\prime$ and having the same sign for neighboring sites
(although for larger distances we may expect some oscillations),
their sign will dominate the overall sign of the results. We will thus get a
negative change in the IPR not only from interaction between electrons in
the same level but also when the interacting electrons are in
different levels. Hence, $\Delta P_n^{-1}$ will always be
negative, as can be seen in the numerical results.

Moreover, repeating the calculations with the non-universal correction to the
averages of wave functions product, we can estimate the dependence of the
effect on the system parameters. We expect the total change in the IPR of a
spin-up electron due to its interaction with $n_\downarrow$ spin-down
electrons to vary as
\begin{eqnarray} \label{eqn:estim}
\Delta P_n^{-1} \sim - \frac{1}{g} \frac{U_H}{\Delta}
\frac{n_\downarrow}{N^3}.
\end{eqnarray}
This expression does not include a factor coming from the sum over
energy denominators, which has only a weak dependence on $N$ and
$n_\downarrow$ (logarithmic for equidistant levels, a weak power
law for a non-constant density of states). Because wave functions
corresponding to neighboring levels are more correlated than
wave-functions corresponding to far away levels, there is also a
factor, which changes sharply (though
not discontinuously) when we pass from $n \le n_\downarrow$ to
$n>n_\downarrow$, as seen in the numerical results. Since $\Delta
\sim N^{-1}$ in real systems (although not in RMT), the effect is
of order $g^{-1}N^{-1}$, if we keep the concentration of spin-down
electrons constant. This is in contrast to the $N^{-2}$ dependence
in RMT. Because $N/g$ is much larger than unity in our numerical
calculations, we can now understand the order of magnitude
difference between RMT and Anderson model results. Thus, all the
features of the numerical data can be explained by taking
non-universal corrections into account.

As we have mentioned before, the non-universal part of the IPR without
interaction, i.e., the difference between the value of the IPR without
interaction in the Anderson model and its value in RMT, varies as
$g^{-1}N^{-1}$. According to the our estimate, the change in the IPR due to
interaction in the Anderson model also goes as $g^{-1}N^{-1}$. Thus, their
ratio, $\Delta P_n^{-1}/(P^{-1}-P_{RMT}^{-1})$, should be independent of
$g$, i.e. of the degree of disorder. It should also be independent of the
number of lattice sites $N$ if the densities of spin-up and spin-down
electrons are kept constant. Thus, this ratio may be used to test our
conjecture for the parametric form of $\Delta P_n^{-1}$.

We first test the $g$ indepence of the ratio
$\Delta P_n^{-1}/(P^{-1}-P_{RMT}^{-1})$ by
plotting it in Fig.~\ref{fig:an_diprd} for systems with identical lattice
sizes (taken to be $17\times 24$, as in the previous calculations), but
different values of disorder. We can clearly see that the differences between
curves corresponding to different W values are much smaller than the
corresponding differences in Fig.~\ref{fig:an_dipr}. The only exception is the
value W=2.0 (the lowest curve), which shows a marked difference from the other
W values. This is probably due to the fact that for W=2.0 disorder is not high
enough, so the electrons' motion is not fully diffusive, and ballistic boundary
effects may be important.

We now test $N$ independence of the ratio
$\Delta P_n^{-1}/(P^{-1}-P_{RMT}^{-1})$ by plotting
it in Fig.~\ref{fig:an_diprn} for systems with the same value of disorder
(taken as W=4.0) but different lattice sizes -- $8\times 13$, $13\times19$,
$17\times 24$. In all the cases the densities of spin-up and spin-down
electrons are approximately equal (the horizontal axis is not the level number
of the affected spin-up electron as before, but the filling $\nu$, defined as
the ratio of the number of spin-up electrons $n$ and the total number of
lattice sites $N$). We can clearly see that the different
curves are almost identical. The only exception is the small $8\times 13$
lattice, whose slighly different behavior can again be attributed to ballistic
boundary effects.

\section{Conclusions}
In conclusion, we have shown how a spin-dependent interaction can
cause delocalization, at least for weak short-range interaction.
Localized electrons highly repulse each other, especially if they
have the same orbital wave function and thus a different spin.
This results in a tendency for interaction-induced delocalization.
The effect on an electron with a given orbital level and spin
direction is stronger if the same orbital level is occupied by an
electron with an opposite spin, and increases with the total
number of electrons with opposite spin. The delocalization is thus
reduced by an in-plane magnetic field. All this is in accordance,
at least qualitatively, with recent experimental findings
\cite{review01} and numerical simulations \cite{berkovits02,dutch03},
regarding the in-plane magnetoresistance.

We have also seen that the main difference in the influence of the
Hubbard interaction between realistic finite $g$ systems and the
RMT stems from exact cancellation of the leading order long
range and short range terms in the former. Thus, while in RMT a state
is correlated only to the same state with an opposite spin (except for
weak anti-correlations with all other states), for finite $g$
correlations between different states lead to a stronger
repulsion between these states resulting in a stronger
delocalization due to the on-site interactions. Nevertheless, the
order of magnitude and parametric dependence of the IPR can be
calculated using RMT, once the non-universal corrections are
properly taken into account.

\begin{acknowledgments}
Financial support from the Israel Science Foundation is gratefully
acknowledged.
\end{acknowledgments}

\end{document}